\documentclass[12pt]{article}

\usepackage{color,tikz}
\usepackage[unicode,bookmarks,bookmarksopen,bookmarksopenlevel=2,colorlinks,linkcolor=blue,citecolor=green]{hyperref}

\usepackage{amsmath,eucal,amssymb}
\usepackage{mathrsfs,graphicx}

\numberwithin{equation}{section}

\def\im{{\mbox{Im}}}

\def\openone{\leavevmode\hbox{\small1\kern-3.3pt\normalsize1}}

\def\bbbc{{\Bbb C}}
\def\bbbr{{\Bbb R}}

\def\diag{\mbox{diag}\,}

\begin{document}

\begin{center}
{\Large \bf On the $3$-wave  Equations with Constant Boundary Conditions}

\bigskip

{\bf Vladimir S. Gerdjikov and Georgi G. Grahovski}

\medskip

{\it Institute of Nuclear Research and Nuclear Energy,  Bulgarian Academy of Sciences, \\ 72 Tsarigradsko chausee, Sofia 1784, Bulgaria }

\medskip

{\small E-mails: gerjikov@inrne.bas.bg, $\quad$ grah@inrne.bas.bg}

\end{center}

\begin{abstract}
\noindent The inverse scattering transform for a  special case of the 3-wave resonant interaction equations with non-vanishing boundary conditions is studied. The Jost solutions and the fundamental analytic solutions (FAS) for the associated spectral problem are constructed. The inverse scattering problem
for the Lax operator is formulated as a Riemann-Hilbert problem on a Riemannian surface.
The spectral properties of the Lax operator are formulated.
\end{abstract}


\section{Introduction}

One of the important nonlinear models with numerous applications in
physics that appeared at the early stages of development of the inverse
scattering method (ISM), see \cite{ZM,K,1,Manak1,FaTa,KRB}, is the
3-wave resonant interaction model described by the equations:
\begin{eqnarray}\label{eq:*1}
{\rm i}{\partial q_1 \over \partial t} + {\rm i}v_1 {\partial q_1 \over \partial x}+
\varkappa q^*_2 q_3&=&0, \nonumber\\
{\rm i}{\partial q_2 \over \partial t} + {\rm i}v_2 {\partial q_2 \over \partial x} +
\varkappa q^*_1 q_3&=&0, \\
{\rm i}{\partial q_3 \over \partial t} + {\rm i}v_3 {\partial q_3 \over \partial x} +
\varkappa q_1 q_2&=&0. \nonumber
\end{eqnarray}
Here $q_i=q_i(x,t), \, i=1,2,3$,  $\varkappa  $ is the
interaction constant,  $v_i $ are the group velocities of the model and the asterisk stays for complex conjugation.
The 3-wave equations can be solved through the ISM due to the fact that
Eq.~(\ref{eq:*1}) allows a Lax representation (see eq.~(\ref{eq:2.1}) below).
The main result of the pioneer papers \cite{ZM} consist also in
proving that if $u_k(x,t) $, $k=1,2,3 $ satisfy the system (\ref{eq:*1}),
then the one-parameter family of ordinary differential operators (\ref{eq:2.1}) is iso-spectral (assuming vanishing boundary conditions as $|x|\to \infty$: $\lim_{|x|\to \infty}q_k(x,t)=0$, $k=1,2,3$).

The 3- and $N$-wave interaction models describe a special class of wave-wave interactions that are not sensitive on
the physical nature of the waves and bear an universal character. This explains why they find numerous applications
in physics and attract the attention of the scientific community over the last few decades
\cite{ZM,1,Chiu,K,Kaup,KRB,CaDe2,Ca*89,Deg,popa,Dokt,geko,BC,BS}.

The interpretation of the inverse scattering method (ISM) as a generalized Fourier
transform and the expansions over the so called "squared solutions" started in \cite{AKNS} for the
nonlinear Schr\"odinger type equations, was soon
generalized also for the $N$-wave equations \cite{G*86}, see  \cite{gvy08} and the numerous references therein.
It allows one to study all  fundamental properties of the relevant nonlinear evolutionary equations  (NLEE's) which include:
\begin{enumerate}
\item the description of the whole class of NLEE related to a given spectral problem (Lax operator $L(\lambda)$ in the form (\ref{eq:2.1})) solvable by the ISM;

\item derivation of the infinite family of integrals of motion associated with $L(\lambda)$;

\item the Hamiltonian properties of the NLEE's.

\end{enumerate}
For the case of (\ref{eq:*1}) one can show that the model equations
are Hamiltonian \cite{ZM,1} and possess a hierarchy of pairwise compatible
Hamiltonian structures \cite{G*86,gvy08}.
The (canonical) Hamiltonian  of (\ref{eq:*1}) is given by \cite{1,vgrn,K,KRB}:
\begin{eqnarray}\label{eq:1.5}
H_{\rm 3-w} ={1 \over 2}\int_{-\infty }^{\infty }  dx\, \left(  \sum_{k=1}^{3} v_k\left(  q_k \frac{\partial q_k^*}{ \partial x }
- q_k^* \frac{\partial q_k}{ \partial x }\right) +\varkappa (q_3 q_1^* q_2^* + q_3^*q_1q_2 ) \right).
\end{eqnarray}
Along with the case of (\ref{eq:*1}) with vanishing boundary conditions, a special interest deserves the case when some (or all) of the functions $q_k(x,t)$ tend to a constant as $x \to\pm \infty$. Below we choose:
\[
q_{1,2}(x,t)\rightarrow 0, \qquad q_3(x,t)\rightarrow \rho {\rm e}^{{\rm i}\phi_\pm}, \qquad x \to\pm \infty.
\]
Here the constants $\theta =\phi_+ - \phi_- $ and $\rho  $ are of a physical origin \cite{FaTa} and
play a basic role in determining the properties of (\ref{eq:*1}) with constant boundary conditions and its
soliton solutions. More specifically, $\rho  $ characterizes the end-points of the continuous spectrum of (\ref{eq:*1}) of the Lax operator $L(\lambda) $. The discrete
spectrum, in this case, may consist of real simple eigenvalues $\lambda
_k$, $k=1,\dots , N $ lying in the lacuna $-2\rho < \lambda_k < 2\rho
$. To them, there correspond the so-called ``dark solitons'' whose
properties and behavior substantially differ from the ones of the bright
solitons. The dark solitons for the nonlinear Schr\"odinger type equations and their generalizations  with non-vanishing boundary conditions are studied in \cite{gk78,gk83,Leon,KonVek1,ABP2,PBT}. Similar results for the discrete nonlinear Schr\"odinger type equations (the Ablowitz-Ladik hierarchy) are obtained in \cite{ABP,KonVek2}.

It is normal to expect that the properties 1 -- 3, known for the case of vanishing boundary conditions
will have their counterparts for the case of constant boundary conditions. However there is no easy and
direct way to do so. The most concise and systematic treatment of both
problems (on the example of nonlinear Schr\"odinger equation) is given in \cite{FaTa}. It  is shown there that one may relate both cases by taking a limit $\rho \to 0 $. Of course, in this limit most of the
difficulties, related mostly with the end-points of the continuous
spectrum disappear. From \cite{FaTa} one can see that the spectral data, the
analyticity properties of the Jost solutions and the corresponding
Riemann-Hilbert \cite{Gakhov,Vekua,Sh} problem are substantially different and more difficult for
$\rho >0 $.

The aim of the present paper is to study the direct scattering problem for the Lax operator and its spectral properties.
In Section 2 we start with the Lax representation and the construction of the Jost solutions of the
Lax operator $L$. In Section 3 we outline the construction of the fundamental analytic solutions (FAS) of $L$.
We also formulate the Riemann-Hilbert problem
on the relevant Riemannian surface, that is satisfied by the FAS. In section 4 we derive the time
evolution for the scattering matrix. Section 5 is devoted to constructing the resolvent of $L$ in terms of
the FAS and the spectral properties of $L$. The effects of the boundary conditions on the  conserved quantities of the 3-wave
equations are analyzed in Section 6. We finish with a brief discussion and conclusions.

\section{Lax representation and Jost solutions}
The idea of the Inverse Scattering Method (ISM) is based on the possibility to linearise the nonlinear evolutionary equation (NLEE) \cite{2,CaDe,1}. To this end we consider the solution of the NLEE $u_k(x,t)$, $k=1,2,3$ as a potential in the Lax operator $L(\lambda)$.

Consider the pair of Lax operators:
\begin{eqnarray}\label{eq:2.1}
L\psi &\equiv & \left( i{\partial  \over \partial x } +[J,Q(x,t)] -\lambda J \right)\psi
(x,t,\lambda )=0,\\
M\psi &\equiv &\left( i{\partial  \over \partial t } +[I,Q(x,t)] -\lambda I \right)\psi
(x,t,\lambda )=0, \nonumber
\end{eqnarray}
with
\begin{equation}\label{eq:*2}
Q = \left( \begin{array}{ccc} 0 & q_{1} & q_{3} \\ q_{1}^* & 0 & q_{2} \\ q_{3}^* & q_{2}^* & 0 \end{array} \right),
\qquad \begin{aligned}
J &= \diag ( J_1, J_2 , J_3) , \\ I &= \diag ( I_1 , I_2 , I_3).
\end{aligned}\end{equation}
We presume that the potential matrices $Q(x,t)$, $I$ and $J$ are traceless (i.e. the Lax operators take values in the algebra $sl(3, {\Bbb C})$) and the eigenvalues of $I$ and $J$ are ordered as follows: $J_1>J_2>J_3$, $I_1>I_2>I_3$ ($J_1+J_2+J_3=0$ and $I_1+I_2+I_3=0$). Here $\lambda \in {\Bbb C}$ is a spectral parameter.

The compatibility condition for (\ref{eq:2.1}) leads to
\begin{equation}\label{Nw-gen}
i[J,Q_t] - i[I,Q_x] +[[I,Q],[J,Q]] = 0,
\end{equation}
which is equivalent to (\ref{eq:*1}), if the potential matrices are taken from (\ref{eq:*2}). The group velocities in (\ref{eq:*1}) take the form:
\[
v_1 = {I_1-I_2\over J_1-J_2}, \qquad v_2 = {I_2-I_3\over J_2-J_3}, \qquad v_3 = {I_1-I_3\over J_1-J_3},
\]
while the interaction constant $\varkappa$ reads:
\[
\varkappa = J_1I_2+J_2I_3+J_3I_1-J_2I_1-J_3I_2-J_1I_3.
\]
We will assume also, that the potential of the Lax operator is a subject of constant boundary conditions  as $|x|\to \infty$:
\begin{eqnarray}\label{eq:2.2}
\lim_{x\to\pm\infty }q_1(x,t) =\lim_{x\to\pm\infty }q_2(x,t)  = 0, \qquad \lim_{x\to\pm\infty }q_3(x,t) = q_3^\pm=\rho e ^{i\phi_\pm}.
\end{eqnarray}
Equivalently, for the potential matrix $Q(x,t)$ one can write:
\begin{eqnarray*}\label{eq:2.2a}
\lim_{x\to\pm\infty }Q(x,t) =Q_\pm , \qquad Q_\pm = \left( \begin{array}{ccc} 0  & 0 & \rho e ^{i\phi_\pm} \\ 0 & 0 & 0 \\ \rho e ^{-i\phi_\pm} &
0 & 0 \end{array} \right)
\end{eqnarray*}
The difference $\theta = \phi_+-\phi_-$ of the asymptotic phases $\phi_\pm$ plays a crucial r\^ole in the Hamiltonian formulation of the 3-wave model with constant boundary conditions: its values label the leaf on the phase space $\mathcal{M}$ of the model (\ref{eq:*1}) where one can determine the class of admissible functionals, and to construct a Hamiltonian formulation. The two asymptotic potentials $Q_\pm$ are related by
\begin{eqnarray}\label{eq:2.2b}
 Q_+  = Q(\theta)Q_-(t)Q^{-1}(\theta),
\end{eqnarray}
where $\theta = \phi_+-\phi_-$ and
\begin{eqnarray}\label{eq:2.2c}
Q (\theta)= \left( \begin{array}{ccc} e ^{i\theta/2}  & 0 & 0 \\ 0 & 1 & 0 \\ 0  &
0 & e ^{-i\theta/2} \end{array} \right).
\end{eqnarray}
The direct and the inverse scattering problem for the Lax operator
(\ref{eq:2.2}) will be done for fixed $t $ and in most of the
corresponding formulae $t $ will be omitted.

The starting point in developing the direct scattering transform for the Lax operator (\ref{eq:2.1}) are the eigenfunctions (the so-called Jost solutions) of the auxiliary spectral problem
\begin{equation}\label{eq:Jost_x}\begin{split}
L(x,t,\lambda)\psi_\pm (x,t,\lambda)=0,
\end{split}\end{equation}
determined uniquely by their asymptotic behavior for $x\to\pm\infty$ respectively:
\begin{equation}\label{eq:Joso}\begin{split}
\lim_{x\to\pm\infty} \psi_\pm (x,t,\lambda) e^{iJ(\lambda)x} =\psi_{\pm,0} (\lambda) P(\lambda),
\end{split}\end{equation}
where $P(\lambda)$ is a projector:
\begin{equation}\label{eq:Pla}\begin{split}
P(\lambda)=\diag (\theta (|{\rm Re}\, \lambda|-2\rho), 1, \theta (|{\rm Re}\, \lambda|-2\rho))
\end{split}\end{equation}
and $\theta(z)$ is the step function. As we shall see below, $P(\lambda)$ ensures that the
continuous spectrum of $L$ has multiplicity 3 for $|{\rm Re}\, \lambda|-2\rho >0$ and multiplicity 1,
for $-2\rho < {\rm Re}\, \lambda <2\rho$.
The $x$ and $t$-independent matrices $\psi_{\pm,0} (\lambda)$ in (\ref{eq:Joso}) diagonalize the
asymptotic Lax operators:
\begin{eqnarray}\label{eq:L-as}
L_\pm(x,t,\lambda)=  i{\partial  \over \partial x } +[J,Q_\pm] -\lambda J.
\end{eqnarray}
Indeed,
\begin{equation}\label{eq:Lpm}\begin{split}
([J,Q_\pm] -\lambda J)\psi_{\pm,0} (\lambda) &= -\psi_{\pm,0} (\lambda) J(\lambda),
\end{split}\end{equation}
where
\begin{equation}\label{eq:psipm}\begin{aligned}
J(\lambda)&= -\mbox{diag}\, (J_1 (\lambda), J_2 (\lambda), J_3 (\lambda)),\\
J_1 (\lambda)&={1\over 2}\left[J_2\lambda + (J_1-J_3)\sqrt{\lambda^2-4\rho^2}\right], &\qquad J_2 (\lambda) &=-\lambda J_2, \\
J_3 (\lambda)&={1\over 2}\left[J_2\lambda - (J_1-J_3)\sqrt{\lambda^2-4\rho^2}\right], &\qquad k(\lambda)&=\sqrt{\lambda^2-4\rho^2}.
\end{aligned}\end{equation}
For the choice of $Q_\pm$ as in (\ref{eq:2.2b}) we have:
\begin{equation}\label{eq:E-rho}
\psi_{\pm, 0}(\lambda)= \frac{ 1}{\sqrt{2\lambda (k+\lambda)}}
\left( \begin{array}{ccc} 2\rho & 0 & -(\lambda + k)e ^{i\phi_\pm} \\
0 & 1 & 0 \\
(\lambda + k)e ^{-i\phi_\pm} & 0 & 2\rho
 \end{array}\right)
\end{equation}
Here and below we will deal with the Riemannian surface related to $k(\lambda)$; its first sheet
 is fixed up by the condition: ${\rm sign }\, \im k(\lambda)={\rm sign }\,\im \lambda$.
The Jost solutions $\psi_+(x,t,\lambda)$ and $\psi_-(x,t,\lambda)$ are related by the scattering matrix $T(t,\lambda)$:
\begin{eqnarray}\label{eq:Scat_m}
 T(t,\lambda) = \psi _+^{-1}(x,t,\lambda )\psi _-(x,t,\lambda ), \qquad \det T(\lambda)=1.
\end{eqnarray}
For a sake of convenience, from now on, instead of the spectral parameter $\lambda$ we will be using the so-called "uniformizing variable"
\begin{eqnarray}\label{eq:zeta}
 \zeta  = {1\over 2 \rho}(\lambda + k(\lambda)).
\end{eqnarray}
In terms of $\zeta$ we have:
\[
\lambda =\rho\left(\zeta + {1\over \zeta} \right), \qquad k(\lambda) =\rho\left(\zeta - {1\over \zeta} \right).
\]
Then, the formulas (\ref{eq:psipm}) take the form:
\begin{equation}\label{eq:psipm0}\begin{aligned}
J (\zeta)&=-\rho \,\diag \left( J_3\zeta + {J_1\over \zeta} , \;  J_2 \left(\zeta + {1\over \zeta}\right) , \;
J_1\zeta + {J_3\over \zeta} \right).
\end{aligned}\end{equation}
Along with the Jost solutions (\ref{eq:Joso}) it is convenient to consider slightly modified Jost solutions:
\begin{eqnarray}\label{eq:Jost-m}
\eta_\pm(x,\zeta)=\psi^{-1}_{\pm, 0}(\zeta)\psi_\pm (x, \zeta)e ^{iJ(\zeta)x},\quad \lim_{x \to \pm \infty}\eta_\pm(x,\zeta)=\openone,
\end{eqnarray}
and satisfying the following associated  to (\ref{eq:Jost_x}) equation:
\begin{eqnarray}\label{eq:Jost_mx}
\qquad {i}\frac{\partial \eta_\pm}{ \partial x } +\psi^{-1}_{\pm, 0}[J,Q(x,t)-Q_\pm]\psi_{\pm, 0}\eta_\pm  -\rho (\zeta +\zeta^{-1})
 [J(\zeta), \eta _\pm(x,t,\zeta )] =0.
\end{eqnarray}
Equivalently, $\eta_\pm(x,\zeta)$ can be regarded as solutions to the following Volterra-type integral equations:
\begin{multline}\label{eq:Jost_int}
\eta_\pm(x,\zeta)=\openone +i\int_{\pm \infty}^{x}d y\,e ^{-iJ(\zeta)(x-y)} \\
\times \psi^{-1}_{\pm, 0}[J,Q(y)-Q_\pm]\psi_{\pm, 0}\eta_\pm(y,\zeta)e ^{iJ(\zeta)(x-y)}, \qquad \zeta\in \bbbr,
\end{multline}
where the  diagonal matrix $J(\zeta)$ is given in (\ref{eq:psipm0}).

In addition the second column of the Jost solution  is defined also on the unit circle in the $\zeta$-plane:
\begin{multline}\label{eq:Jost_2}
(\eta^{(2)}_\pm)_{k2}(x,\zeta)= \delta_{k2} + i\int_{\pm \infty}^{x}d y\,e ^{-i(J_k(\zeta)-J_2(\zeta))(x-y)} \\
\times \left(\psi^{-1}_{\pm, 0}[J,Q(y)-Q_\pm]\psi_{\pm, 0}\eta_\pm(y,\zeta)\right), \qquad |\zeta|=1.
\end{multline}

\section{The fundamental analytic solutions of $L$}
In order to construct the fundamental analytic solutions (FAS) of $L$ we first need to determine
the regions of the complex $\zeta$-plane in which the imaginary parts of the
eigenvalues of $J(\zeta)$ are ordered. To do this we first need to find the curves on which
\begin{equation}\label{eq:ImJ'}\begin{split}
\im (J_j(\zeta) -J_k(\zeta) )=0, \qquad 1\leq j< k \leq 3; \qquad \im J_1(\zeta)=0.
\end{split}\end{equation}
Writing down $\zeta =|\zeta|e^{i\phi_0}$ we find:
\begin{equation}\label{eq:J_k}\begin{split}
\im (J_1(\zeta) -J_2(\zeta)) &=\rho \left( (J_2-J_3)|\zeta| + \frac{J_1 -J_2}{|\zeta|} \right) \sin \phi_0, \\
\im (J_2(\zeta) -J_3(\zeta)) &=\rho \left( (J_1-J_2)|\zeta| + \frac{J_2 -J_3}{|\zeta|} \right) \sin \phi_0, \\
\im (J_1(\zeta) -J_3(\zeta)) &=\rho  (J_1-J_3) \left(|\zeta| + \frac{1}{|\zeta|} \right) \sin \phi_0,\\
\im J_2(\zeta) &=-J_2 \rho \left(|\zeta| - \frac{1}{|\zeta|}\right) \sin \phi_0.
\end{split}\end{equation}
Since $J_1-J_2>0$, $J_1-J_3>0$ and $J_2-J_3>0$ it is easy to see that the  solutions of the eqs. (\ref{eq:ImJ'})
are $\phi_0=0$ and $\phi_0=\pi$, i.e. the real axis in the complex $\zeta$-plane; in addition $\im J_2(\zeta)=0$
for $|\zeta|^2=1$.

\begin{figure}[t]
  \includegraphics[width=6.5cm]{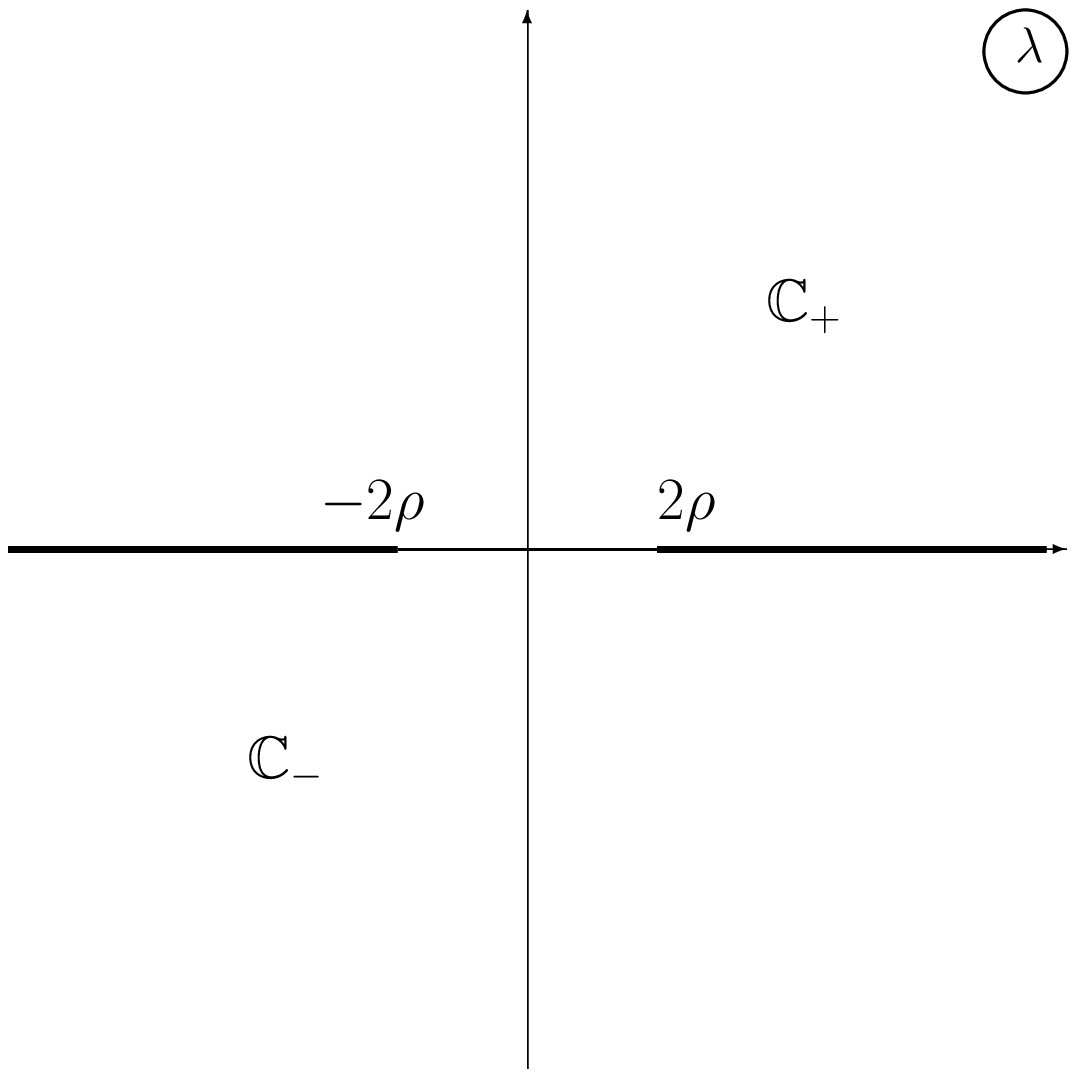} \quad   \includegraphics[width=6.5cm]{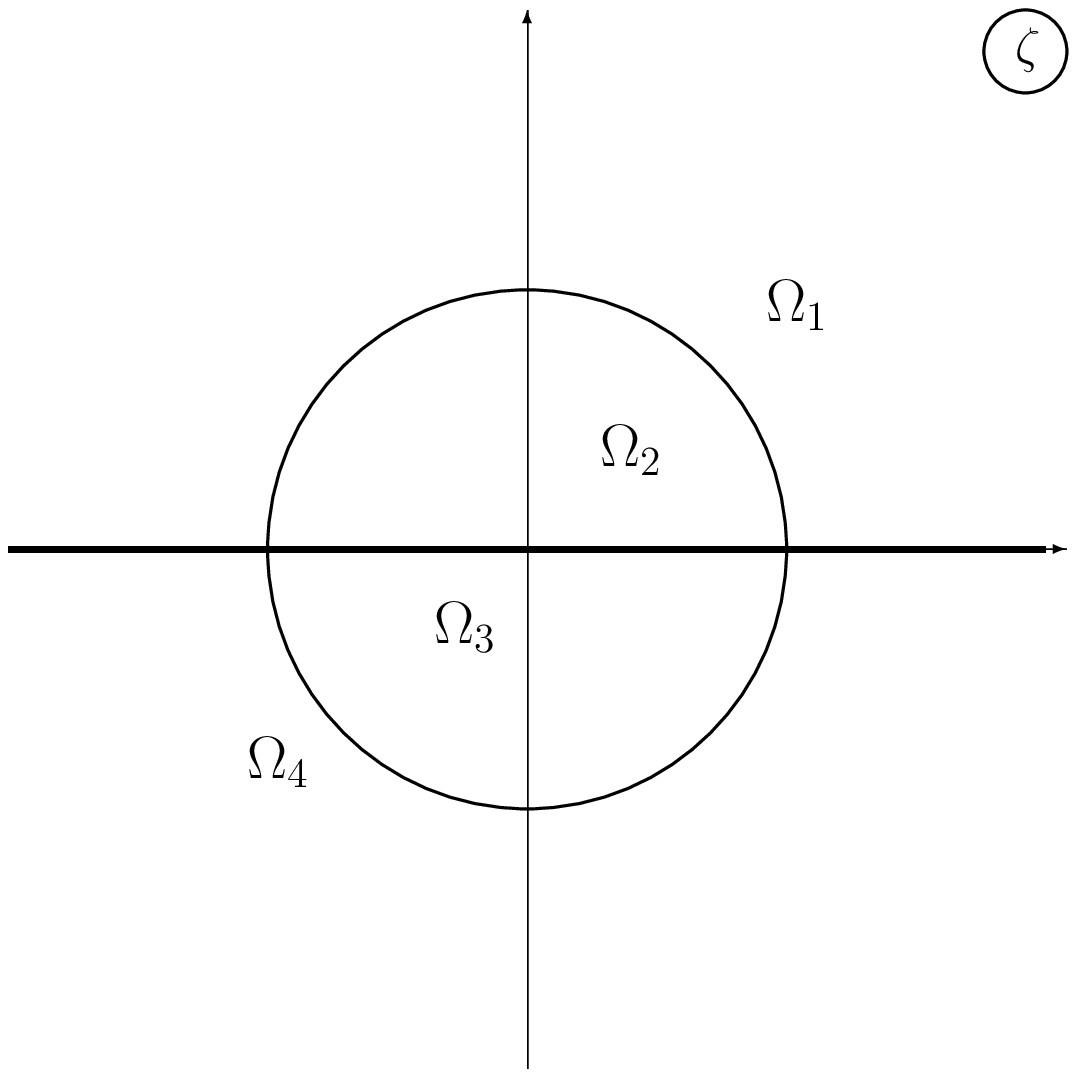}
  \caption{The continuous spectrum of $L$ in the complex $\lambda$ plane (left panel) and
in the complex $\zeta$ plane (right panel),  \label{fig:1}}
\end{figure}

The complex $\zeta$-plane is split into four regions $\Omega_k$, $k=1,\dots,4$ formed by the intersections
of the upper and lower complex half-planes $\bbbc_+$ and $\bbbc_-$ with the unit circle $\mathcal{S}$,
see figure \ref{fig:1}. The ordering of $\im J_k(\lambda)$ in each of them  depends on the sign of $J_2$
and are as follows:
\begin{equation}\label{eq:Jk-ord}\begin{aligned}
&\Omega_1: & \qquad  \im J_1(\lambda)> 0 > \im J_2(\lambda) >\im J_3(\lambda), \\
&\Omega_2: & \qquad  \im J_1(\lambda)> \im J_2(\lambda) > 0  >\im J_3(\lambda), \\
&\Omega_3: & \qquad  \im J_3(\lambda)> 0 >\im J_2(\lambda) >\im J_1(\lambda),\\
&\Omega_4: & \qquad  \im J_3(\lambda)> \im J_2(\lambda) > 0  >\im J_1(\lambda),
\end{aligned}\end{equation}
for $J_2>0$ and
\begin{equation}\label{eq:Jk-ord2}\begin{aligned}
&\Omega_1: & \qquad  \im J_1(\lambda)> \im J_2(\lambda)> 0  >\im J_3(\lambda), \\
&\Omega_2: & \qquad  \im J_1(\lambda)> 0  >\im J_2(\lambda)  >\im J_3(\lambda), \\
&\Omega_3: & \qquad  \im J_3(\lambda)> \im J_2(\lambda) > 0 >\im J_1(\lambda),\\
&\Omega_4: & \qquad  \im J_3(\lambda)>  0  >\im J_2(\lambda)  >\im J_1(\lambda),
\end{aligned}\end{equation}
for $J_2<0$.

Let us first construct the FAS in the region $\Omega_1 $. Following the ideas of \cite{1,gvy08} we introduce it
as the solution of the following set of integral equations:
\begin{multline}\label{eq:xip}
\left\{\xi_{(1)}^+(x,\zeta)\right\}_{kl}=\delta_{kl} +{ i}\int_{\infty}^{x}{ d}y\,{ e}^{- i(J_k(\zeta)-J_l(\zeta))(x-y)}\\
\times \left\{\psi_{+, 0}^{-1}[J,Q(y)-Q_+]\psi_{+, 0}\xi_{(1)}^+(y,\zeta)\right\}_{kl},\qquad k< l.
\end{multline}
\vspace{-15pt}
\begin{multline*}
\left\{\xi_{(1)}^+(x,\zeta)\right\}_{kl}= { i} \int^{x}_{-\infty}{ d}y\, e^{-i (J_{k}(\zeta)  -J_{ l}(\zeta))(x-y)}\\
 \times\left\{\psi_{+, 0}^{-1}[J,Q(y)-Q_-]\psi_{+, 0}\xi_{(1)}^+(y,\zeta\right\}_{kl}, \qquad k \geq l.
\end{multline*}
The proof of the fact that $\xi_{(1)}^+ (x,\zeta)$ is an analytic function of $\zeta$ for any $\zeta\in\Omega_1 $
is based on the fact, that due the ordering (\ref{eq:Jk-ord})  all exponential factors in eqs. (\ref{eq:xip})
for $\zeta\in\bbbc_+ $ are decaying.
This ensures the convergence of all integrals as well as the existence of $\xi_{(1)}^+ (x,\zeta)$ for all $\zeta\in\Omega_1$.
Similar conclusions can be drawn also for the $(d/d\zeta)^k \xi_{(1)}^+ $. Indeed, taking  derivatives of (\ref{eq:xip})
with respect to $\zeta$ would give rise to terms polynomial in $x$ in the integrands. Such terms, however, are
suppressed by the exponential factors, which allows to conclude, that along with $\xi_{(1)}^+ (x,\zeta)$ also
its derivatives $(d/d\zeta)^k \xi_{(1)}^+ $ exist for any positive $k$. Thus, we briefly outlined the idea of the proof
that $\chi_{(1)}^+ (x,\zeta)= \xi_{(1)}^+ (x,\zeta)e^{i J(\zeta)x}$ are FAS of the Lax operator $L$ for $\zeta\in\Omega_1$.

Similarly, the FAS in the region $\Omega_4$ is the solution of the set of integral equations:
\begin{multline}\label{eq:xim}
\left\{\xi_{(4)}^- (x,\zeta)\right\}_{kl}=\delta_{kl} +{ i}\int_{-\infty}^{x}{ d}y\,{ e}^{-i(J_k(\zeta)-J_l(\zeta))(x-y)}\\
\times \left\{\psi_{+, 0}^{-1}[J,Q(y)-Q_-]\psi_{+, 0}\xi_{(4)}^- (y,\zeta)\right\}_{kl},\qquad k\leq l.
\end{multline}
\vspace{-15pt}
\begin{multline*}
\left\{\xi_{(4)}^- (x,\zeta)\right\}_{kl}= { i} \int^{x}_{\infty}{ d}y\, e^{ -i (J_{k}(\zeta)  - J_{ l}(\zeta))(x-y)}\\
 \times\left\{\psi_{+, 0}^{-1}[J,Q(y)-Q_+]\psi_{+, 0}\xi_{(4)}^- (y,\zeta\right\}_{kl}. \qquad k>l.
\end{multline*}
The proof of the analyticity of $\xi_{(4)}^- (x,\zeta)$ for any $\zeta\in\Omega_4  $ is similar to the one for $\xi_{(1)}^+ (x,\zeta)$.

It remains to outline the construction of $ \xi^-_{(2)}(x,\lambda)$ and $\xi^+_{(3)}(x,\lambda)$ for the regions $\Omega_2$ and $\Omega_3$.
To this end we make use of the involution of the Lax operator $L$ that is a consequence of $Q=-Q^\dag$. Then we conclude:
\begin{equation}\label{eq:Inv}\begin{aligned}
\chi_{(3)}^-(x,\zeta) &= ( \chi_{(1)}^{+,\dag})^{-1}(x,1/\zeta^*), \qquad
\chi_{(2)}^+(x,\zeta) &= ( \chi_{(4)}^{-,\dag})^{-1}(x,1/\zeta^*).
\end{aligned}\end{equation}
The next step is to analyze the interrelations between the Jost solutions $\psi_\pm (x,\zeta)$ and the FAS $\chi^+ (x,\zeta)$
and  $\chi^- (x,\zeta)$. It is natural to expect that they are linearly related. Skipping the details we note that:
\begin{equation}\label{eq:chi=psi}\begin{aligned}
\chi^+_{(\alpha)} (x,\zeta) & = \psi_-(x,\zeta) S^+_{(\alpha)}(\zeta), &\qquad \chi^+_{(\alpha)}
 (x,\zeta) & = \psi_+(x,\zeta)T^-_{(\alpha)}(\zeta) D^+_{(\alpha)}(\zeta),\\
\chi^-_{(\beta)} (x,\zeta) & = \psi_-(x,\zeta) S^-_{(\beta)}(\zeta), &\qquad \chi^-_{(\beta)} (x,\zeta) & =
\psi_+(x,\zeta)T^+_{(\beta)}(\zeta) D^-_{(\beta)}(\zeta),
\end{aligned}\end{equation}
where $\zeta\in\bbbr\cup \mathcal{S}$, $\alpha=1,3$ and $\beta=2,4$
match the indices of the regions of analyticity $\Omega_k$. We will often omit the indices $\alpha$ and
$\beta$, since their values are clear from the figure \ref{fig:1}.
Here $S^+_{(\alpha)}$ and $T^+_{(\alpha)}$ (resp.  $S^-_{(\beta)}$ and $T^-_{(\beta)}$) are upper triangular
(resp. lower triangular) matrices whose diagonal elements are all
equal to 1; the matrices $D^\pm_{(\alpha)}$, $D^\pm_{(\beta)}$  are diagonal ones. In fact these matrices
are directly related to the Gauss decomposition of the scattering matrix $T(\zeta)$ (\ref{eq:Scat_m}):
\begin{equation}\label{eq:T}
T(\zeta)=T^{\mp}(\zeta)D^{\pm}(\zeta)(S^{\pm}(\zeta))^{-1},
\end{equation}
where:
\begin{equation}\label{eq:Gauss}\begin{aligned}
 T^+(\zeta)&=\left(\begin{array}{ccc} 1 & T_1^+(\zeta) & T_3^+(\zeta) \\ 0 & 1 & T_2^+(\zeta)\\
0 & 0 & 1\end{array} \right), &\quad  T^-(t,\zeta)&=\left(\begin{array}{ccc} 1 & 0 & 0 \\ T_1^-(\zeta) & 1 & 0\\
T_3^-(\zeta) & T_2(\zeta) & 1\end{array} \right), \\
S^+(\zeta) &=\left(\begin{array}{ccc} 1 & S_1^+(\zeta) & S_3^+(\zeta) \\ 0 & 1 & S_2^+(\zeta)\\
0 & 0 & 1\end{array} \right), &\quad  S^-(t,\zeta) &=\left(\begin{array}{ccc} 1 & 0 & 0 \\ S_1^-(\zeta) & 1 & 0\\
S_3^-(\zeta) & S_2(\zeta) & 1\end{array} \right),
\end{aligned}\end{equation}
\begin{equation}\label{eq:Dpm}\begin{split}
D^+(\zeta) &=\mbox{diag}\, \left(m_1^+(\zeta), \frac{ m_2^+(\zeta)}{m_1^+(\zeta)}, \frac{1}{m_2^+(\zeta)} \right), \\
D^-(\zeta) &=\mbox{diag}\, \left( \frac{1}{m_2^-(\zeta)}, \frac{ m_2^-(\zeta)}{m_1^-(\zeta)}, m_1^-(\zeta) \right),
\end{split}\end{equation}
Here $m_k^\pm(\zeta)$ are the principal upper/lower minors of order $k$ of the scattering matrix (\ref{eq:Scat_m}); the explicit expressions
for the matrix elements of $T^\pm$ and $S^\pm$ in terms of $T_{ij}(\zeta)$ are given by:
\begin{eqnarray}\label{eq:Gauss-f}
S_1^+(\zeta)=-{T_{12}(\zeta)\over T_{11}(\zeta)},\quad S_2^+(\zeta)=  {T_{13}(\zeta) T_{21}(\zeta)- T_{11}(\zeta) T_{23}(\zeta)\over T_{11}(\zeta)T_{22}(\zeta) - T_{12}(\zeta)T_{21}(\zeta)},\nonumber\\
S_3^+(\zeta)=  {T_{12}(\zeta) T_{23}(\zeta)- T_{13}(\zeta) T_{22}(\zeta)\over T_{11}(\zeta)T_{22}(\zeta) - T_{12}(\zeta)T_{21}(\zeta)},\\
T_1^-(\zeta)={T_{21}(\zeta)\over T_{11}(\zeta)},\quad T_2^-(\zeta)=  {T_{32}(\zeta) T_{11}(\zeta)- T_{31}(\zeta) T_{12}(\zeta)\over T_{11}(\zeta)T_{22}(\zeta) - T_{12}(\zeta)T_{21}(\zeta)},\nonumber\\
T_3^-(\zeta)=  {T_{31}(\zeta)\over T_{11}(\zeta)},
\end{eqnarray}
and
\begin{eqnarray}\label{eq:Gauss-f2}
S_1^-(\zeta)={T_{33}(\zeta) T_{21}(\zeta)- T_{23}(\zeta) T_{31}(\zeta)\over T_{32}(\zeta)T_{23}(\zeta) - T_{33}(\zeta)T_{22}(\zeta)},\\
S_2^-(\zeta)=  -{T_{32}(\zeta)\over T_{33}(\zeta)}, \qquad
S_3^-(\zeta)= {T_{31}(\zeta) T_{22}(\zeta)- T_{32}(\zeta) T_{21}(\zeta)\over T_{32}(\zeta)T_{23}(\zeta) - T_{33}(\zeta)T_{22}(\zeta)},\nonumber\\
T_1^-(\zeta)={T_{12}(\zeta) T_{33}(\zeta)- T_{13}(\zeta) T_{32}(\zeta)\over T_{11}(\zeta)T_{33}(\zeta) - T_{13}(\zeta)T_{31}(\zeta)},\\
 T_2^-(\zeta)= {T_{23}(\zeta)\over T_{33}(\zeta)} ,\qquad
T_3^-(\zeta)=  {T_{13}(\zeta)\over T_{33}(\zeta)}.\nonumber
\end{eqnarray}
Due to the the special choice of the matrix $Q(x,t)$ (\ref{eq:*2}) it follows that $S^-(\zeta)=(S^+(1/\zeta^*))^\dag$
and $T^+(\zeta)=(T^-(1/\zeta^*))^\dag$, so the matrix elements of $S^-(\zeta)$ and $T^+(\zeta)$ can be reconstructed from (\ref{eq:Gauss-f}).

One of the most effective method for solving the inverse scattering problem for a given
Lax operator $L$ is to reduce it to a RHP \cite{Sh}. On the complex $\zeta$-plane it can be formulated  as follows:
\begin{equation}\label{eq:rhp1}\begin{split}
\xi^{+}_{(\alpha)} (x,\zeta) &=\xi^{-}_{(\beta)}(x,\zeta)G_{\alpha,\beta}(x,t,\zeta), \qquad  \lim_{k\to\infty} \xi^{+} (x,\zeta) =\openone, \\
 G(\zeta) &=  e^{-iJ(\zeta)x -iF(\zeta)t}(S^-)^{-1}S^+ e^{iJ(\zeta)x+iF(\zeta)t}.
\end{split}\end{equation}
The relation (\ref{eq:rhp1}) holds true for $k\in \bbbr$ in the complex $k$-plane.
The RHP for Lax operators with vanishing boundary conditions look similarly.
However, the relation (\ref{eq:rhp1}) is  more complicated due to the fact,
that we are dealing with an RHP formulated  on the Riemannian surface related
to the root $k(\lambda)=\sqrt{\lambda^2 -4\rho^2}$.

The sewing function $G(x,\zeta)$ gives the minimal set of scattering data, sufficient to reconstruct the scattering matrix $T(\zeta)$.

One of the important uses of the RHP is that it allows one to use the
Zakharov-Shabat dressing method and construct the soliton solutions of the relevant NLEE.
Doing this one should specify the dressing factor as a rational function of the
uniformizing variable $\zeta$ rather than $\lambda$.

\section{The time evolution of the scattering matrix}
We start by noting that we could use a bit more general $M$-operator than the one in (\ref{eq:2.1}),
namely:
\begin{eqnarray}\label{eq:2.1m}
M\psi &\equiv &\left( i{\partial  \over \partial t } +[I,Q(x,t)] -\zeta I \right)\psi (x,t,\zeta )=\psi (x,t,\zeta ) F(\zeta).
\end{eqnarray}
The compatibility condition $[L,M]=0$ holds true for any $x$- and $t$-independent matrix $F(\zeta)$. We will fix up $F(\zeta)$,
requiring that the definition of the Jost solutions (\ref{eq:Joso}) holds true for all $t$. Let us identify $\psi (x,t,\zeta )$
as $\psi_+ (x,t,\zeta )$ (resp. $\psi_- (x,t,\zeta )$) and take the limit $x\to\infty$ (resp. $x\to -\infty$). This gives:
\begin{equation}\label{eq:psiM}\begin{split}
([I,Q_\pm] - \zeta I)\psi_{\pm,0} = \psi_{\pm,0} F(\zeta).
\end{split}\end{equation}
It is easy to check that $\psi_{\pm,0} $ diagonalize also  $[I,Q_\pm]-\zeta I$ and therefore $F(\zeta)$ is a diagonal matrix:
\[
F(\zeta)=\mbox{diag}\, (f_1(\zeta), f_2(\zeta), f_3(\zeta)),
\]
where $f_j(\zeta)$ are the eigenvalues of $[I,Q_\pm]-\zeta I$. In terms of $\zeta$ we have:
\begin{eqnarray}\label{eq:M-eigvals}
F (\zeta)=\rho \, \diag \left( I_1\zeta + \frac{I_3}{\zeta}, \; I_2 \left(\zeta - \frac{1}{\zeta}\right) , \;
I_3 \zeta + \frac{I_1}{\zeta} \right).
\end{eqnarray}
Next, we insert $\psi =\psi_+ (x,t,\zeta )$ into (\ref{eq:2.1m}) and take the limit $x\to -\infty$. Thus we obtain,
that if the matrix elements $u_k(x,t)$, $k=1,2,3$ of the potential of $L(\zeta)$ satisfy the NLEE (\ref{eq:*1}) then the time evolution of the associated scattering matrix is given by the linear ODE:
\begin{eqnarray}\label{eq:T-t}
i{d T\over d t} -  [F(\zeta),T(t,\zeta)]=0.
 \end{eqnarray}
As a consequence the Gauss factors of $T(t,\zeta)$ satisfy
\begin{equation}\label{eq:dspm}\begin{aligned}
i{d T^\pm \over d t} -  [F(\zeta),T^\pm (t,\zeta)]=0, \quad &\qquad i{d S^\pm \over d t} -  [F(\zeta),S^\pm (t,\zeta)]=0,  \quad
i{d D^\pm \over d t} &=0,
\end{aligned}\end{equation}
From the last equation it follows that the principle minors $m_1^\pm (\zeta)$ and $m_2^\pm (\zeta)$ of the scattering matrix are
time-independent and can be considered as generating functionals of the integrals of motion for (\ref{eq:*1}), while for the off-diagonal ones we get:
\begin{eqnarray}\label{eq:bk-t}
T_{ij}(t,\zeta)= T_{ij}(0,\zeta)e ^{-i(f_i(\zeta)-f_j(\zeta))t}.
 \end{eqnarray}
The function $F(\zeta)$ is known as the dispersion law for the 3-wave equations with constant boundary conditions.

\bigskip

\section{Spectral Properties of the Lax Operator}

The crucial fact that determines the spectral properties of the operator
$L (\zeta)$ is the choice of the class of functions where from we shall choose
the potential $Q(x) $. Below, for a sake of simplicity, we assume that   $Q(x,t) $ satisfies

\medskip
{\bf Condition C.1} $Q(x,t) $  is smooth for all $x$ and $t$ and is such that
\[
\lim_{x\to\pm\infty} |x|^p (Q(x,t) -Q_\pm) =0\quad  \mbox{for all} \quad p=0,1,\dots.
\]
\medskip

The FAS $\chi ^\pm    (x,\zeta ) $ of $L(\zeta ) $ allows one to construct
the resolvent of the operator $L $ and then to investigate its spectral
properties. By a resolvent of $L(\zeta ) $ we understand an integral
operator $R(\zeta ) $ with kernel $R(x,y,\zeta ) $ which satisfies
\begin{equation}\label{eq:R1.1}
L(\zeta )(R(\zeta )f)(x)=f(x),
\end{equation}
where $f(x) $ is an $3 $-component vector complex-valued function  with
bounded norm, i.e. 
\[
\int_{-\infty }^{\infty } dy |f^T(y) f(y)| <\infty .
\]
From the general theory of linear operators  we
know that the point $\zeta $ in the complex $\zeta  $-plane is
a regular point if $R(\zeta ) $ is a bounded integral operator.
In each connected subset of regular points $R(\zeta ) $ is
analytic in $\zeta  $.
The points $\zeta  $ which are not regular constitute the spectrum of
$L(\zeta ) $. Roughly speaking, the spectrum of $L(\zeta ) $ consist of
two types of points:

\begin{itemize}

\item i) the continuous spectrum of $L(\zeta ) $ consists of all points
$\zeta  $ for which $R(\zeta ) $ is an unbounded integral operator;

\item ii) the discrete spectrum of $L(\zeta ) $ consists of all points
$\zeta  $ for which $R(\zeta ) $ develops pole singularities.

\end{itemize}

Let us now show how the resolvent $R(\zeta ) $ can be expressed through
the FAS of $L(\zeta ) $. Indeed, if we write down $R(\zeta ) $ in the
form:
\begin{equation}\label{eq:R1.2}
R(\zeta ) f(x) = \int_{-\infty }^{\infty } R(x,y,\zeta ) f(y),
\end{equation}
the kernel $R(x,y,\zeta ) $ of the resolvent is given by:
\begin{equation}\label{eq:R2.1}
R_{(\alpha)}(x,y,\zeta ) = R_{(\alpha)}^\pm   (x,y,\zeta ), \qquad \mbox{for\;} \zeta \in \Omega_{(\alpha)} .
\end{equation}
where
\begin{equation}\label{eq:R}\begin{aligned}
R_{(\alpha)}^\pm  (x,y,\zeta) &= -i\chi_{(\alpha)}^\pm  (x,\zeta) \Theta_{(\alpha)}^\pm  (x-y) (\chi_{(\alpha)}^\pm  )^{-1}(y,\zeta),
\qquad \zeta\in \Omega_{(\alpha)}  ,
\end{aligned}\end{equation}
and
\begin{equation}\label{eq:R2.2}\begin{split}
\Theta_1 ^+  (x-y) &= \diag (-\theta (y-x), \theta(x-y), \theta (x-y)), \\
 \Theta_2 ^-  (x-y) &= \diag (-\theta (y-x), -\theta(y-x), \theta (x-y)) , \\
\Theta_3 ^+ (x-y) &= \diag (\theta (x-y), -\theta(y-x), -\theta (y-x)), \\
  \Theta_4 ^-  (x-y) &= \diag (-\theta (x-y), \theta(x-y), -\theta (y-x)),
\end{split}\end{equation}
for $J_2>0$, and
\begin{equation}\label{eq:R2.2a}\begin{split}
\Theta_1 ^+  (x-y) &= \diag (-\theta (y-x), -\theta (y-x), \theta (x-y)), \\
 \Theta_2 ^-  (x-y) &= \diag (-\theta (y-x), \theta(x-y), \theta (x-y)) , \\
\Theta_3 ^+ (x-y) &= \diag (\theta (x-y), \theta(x-y), -\theta (y-x)), \\
  \Theta_4 ^-  (x-y) &= \diag (-\theta (x-y), -\theta (y-x), -\theta (y-x)),
\end{split}\end{equation}
for $J_2<0$. The next theorem establishes that $R (x,y,\zeta ) $ is indeed the kernel
of the resolvent of $L(\zeta ) $.

{\bf Theorem 1.} {\it
Let $Q(x) $ satisfy the conditions (C.1) and is such that  the minors $m_k^\pm(\zeta ) $
have a finite number of simple zeroes $\zeta _j^\pm $. Then
\begin{enumerate}

\item $R^\pm   (x,y,\zeta ) $ is an analytic function of $\zeta  $ for
$\zeta \in \bbbc_\pm  $ having pole singularities at $\zeta _j^\pm $;

\item $R^\pm   (x,y,\zeta ) $ is a kernel of a bounded integral operator
for $ \zeta \in \bbbr \cup \mathcal{E}$;

\item $R (x,y,\zeta ) $ is uniformly bounded function for $\zeta
\in\bbbr \cup \mathcal{E} $ and provides a kernel of an unbounded integral operator;

\item $R^\pm   (x,y,\zeta ) $ satisfy the equation:
\begin{equation}\label{eq:R3.1}
L(\zeta ) R^\pm   (x,y,\zeta )=\openone \delta (x-y).
\end{equation}
\end{enumerate}
}

{\bf Proof:}
\begin{enumerate}

\item is obvious from the fact that $\chi ^\pm(x,\zeta ) $ are the FAS
of $L(\zeta ) $;

\item Assume that $ \zeta \in \Omega_1  $ and consider the asymptotic behavior
of $R^+  (x,y,\zeta ) $ for $x,y\to\infty  $. From equations (\ref{eq:chi=psi}) and (\ref{eq:R}) we find that
\begin{eqnarray*}\label{eq:R3.2}
R_{ij}^+ (x,y,\zeta ) &=& \sum_{p=1}^{3} \chi^+_{ip}(x,\zeta )
e^{-i J_p(\zeta) (x-y)} \Theta^+_{1,pp}(x-y) \hat{\chi}^+_{pj}(y,\zeta ) .
\end{eqnarray*}

Due to the fact that $\chi ^+ (x,\zeta ) $ has triangular asymptotics
for $x\to\infty  $ and $\zeta \in\bbbc_+  $ and for the correct choice of
$\Theta^+ (x-y) $ (\ref{eq:R2.2}) we check that the right hand side of
(\ref{eq:R3.2}) falls off exponentially for $x\to\infty  $ and  an arbitrary
choice of $y $. All other possibilities are treated analogously \cite{vsg82,G*86}.

\item For $\zeta \in\bbbr \cup \mathcal{S}$ the arguments of 2) can not be applied
because the exponentials in the right hand side of (\ref{eq:R3.2}) only oscillate. Thus we conclude that
$R^\pm  (x,y,\zeta ) $ for $\zeta \in\bbbr\cup \mathcal{S} $ is only a bounded
function and thus the corresponding operator $R(\zeta ) $ is an
unbounded integral operator.

\item The proof of eq. (\ref{eq:R3.1}) follows from the fact that
$L(\zeta )\chi^\pm  (x,\zeta )=0 $ and
\begin{equation}\label{eq:R4.1}
{d\Theta^\pm (x-y)  \over dx } = \openone \delta (x-y).
\end{equation}
\end{enumerate}
$\blacksquare$

Thus we conclude that the continuous spectrum of $L$ in the complex $k$-plane coincides
with the contour of the RHP $\bbbr\cup \mathcal{S}$ with multiplicity 3 on $\bbbr$ and
multiplicity 1 on $\mathcal{S}$. On the complex $\lambda$-plane the continuous spectrum of
$L$ is on the real axis; it has multiplicity 3 on the semi-axis $ {\rm Re}\, \lambda<-2\rho$ and  $ {\rm Re}\, \lambda> 2\rho$
and multiplicity 1 in the `lacuna'  $ -2\rho < {\rm Re}\, \lambda <2\rho$.

\bigskip

\section{Conserved Quantities for the 3-wave Equations}

As we already mentioned above, the diagonal factors $D^\pm (\zeta)$ are time independent and can be used to generate the infinite set of integrals of motion for (\ref{eq:*1}). For the 3-wave resonant interaction equations, these matrices are expressed through the principal upper/lower minors $m^\pm(\zeta)$ of the scattering matrix $T(\zeta)$ (\ref{eq:Scat_m}).
 Skipping the details (see \cite{G*86}) we get:
\begin{eqnarray}\label{eq:Dj1}
&& \ln D^\pm_{k,1} = -{ i \over 4 } (J_k -  J_{k+1}){\cal P}_k + (J_1-J_3){\cal P}_3 ,
\end{eqnarray}
The momenta ${\cal P}_k$, $k=1,2,3$ are given by:
\begin{equation}\label{eq:M-R-P}
{\cal P}_{1}=\int_{-\infty }^{\infty }dx\,
 |q_{1}(x)|^2,\quad {\cal P}_{2}=\int_{-\infty }^{\infty }dx\,
 |q_{2}(x)|^2, \quad {\cal P}_{3}=\int_{-\infty }^{\infty }dx\,
 (|q_{3}(x)|^2 - \rho^2).
\end{equation}
The fact that $\ln m^\pm_{1} $ generates integrals of motion can be considered as natural analog of the Manley--Rowe relations
\cite{ZM,K}. In the case of (\ref{eq:*2}), then (\ref{eq:Dj1}) is equivalent
to the existence of two additional first integrals for the model (\ref{eq:*1})
\begin{equation}\label{eq:M-R}\begin{split}
I_1&= (J_1-J_2){\cal P}_{1} + (J_1-J_3){\cal P}_{3}=\mbox{const},\\
I_2&= (J_2-J_3){\cal P}_{2} + (J_1-J_3){\cal P}_{3}=\mbox{const},
\end{split}\end{equation}
which are linear combinations of the momenta (\ref{eq:M-R-P})
and can be interpreted as relations between the densities $|q_\alpha |^2 $
of the waves of type $\alpha $. The total momentum for the 3-waves is also a conserved quantity:
\begin{eqnarray}\label{eq:Momentum}
{\cal P}=(J_1-J_2){\cal P}_{1} + (J_2-J_3){\cal P}_{2} + (J_1-J_3){\cal P}_{3} =\mbox{const}.
\end{eqnarray}
The integral of motion $D_{2} $ is proportional to the
Hamiltonian of the $3$-wave equations (\ref{eq:1.5}).

For the case of constant boundary conditions, the functional $\mathcal{P}_3$ is subject of regularization (\ref{eq:Dj1})
by using the asymptotic values of the potential $Q_\pm$, while the functional $H_{\rm 3-w}$ remains the same as
for the case of vanishing boundary conditions.

\bigskip

\section{Conclusions}

We  studied the direct scattering problem for the Lax operator and its spectral properties.
This includes: the construction of  Lax representation and  the Jost solutions of the
Lax operator $L $. Furthermore, we outlined the construction of the fundamental analytic solutions (FAS) of $L$
and  formulated a  Riemann-Hilbert problem for the FAS on a relevant Riemannian surface.

We also outlined the  construction of the resolvent of $L(\zeta)$ in terms of
the FAS and the spectral properties of $L$. Finally, we briefly discuss the effects of the boundary conditions on the
conserved quantities of the 3-wave equations: we showed that the total momentum for non-vanishing boundary conditions
needs regularization, while the Hamiltonian remains the same.

Similar analysis can be done for a $3$-wave resonant interaction model with more general boundary conditions: $\lim_{x\to -\infty} q_k (x,t)= q_k^-$ ($k=1$ or $2$) and $\lim_{x\to +\infty} q_3 (x,t)= q_3^-$. This may require the matrices $Q(\theta)$ (\ref{eq:2.2c}) to have also off-diagonal entries.

It is an open problem to to derive the soliton solutions of (\ref{eq:*1}) in the case of constant boundary conditions (the so-called "dark solitons"), the dark-dark and dark-bright soliton solutions \cite{ABP2} by modifying the dressing Zakharov-Shabat method \cite{1}, or by using the Darboux transformation method \cite{Deg}.

Another challenge is to extend this analysis also for systems, describing resonant interactions of $N$ waves \cite{1,vgn1} or to $N$-wave type systems related to simple Lie algebras \cite{vgn1,vgn2}.

Another open problem is to study the behavior of the scattering data at the end-points of the continuous spectrum in the complex $\lambda$-plane;
this requires generalization of the the method developed in  \cite{FaTa}.

\label{last}
\end{document}